\documentclass[prb, twocolumn, groupedaddress, superscriptaddress, showkeys, preprintnumbers]{revtex4}

\usepackage{graphicx}

\begin{document}

\title{Softening of ultra-nanocrystalline diamond at low grain sizes}

\author{Ioannis N. Remediakis}
\affiliation{Department of Physics, University of Crete, P.O. Box 2208, 71003
Heraklion, Crete, Greece}
\affiliation{Department of Materials Science and
Technology, University of Crete, P.O. Box 2208, 71003 Heraklion, Crete}
\email{remed@materials.uoc.gr}
\author{Georgios Kopidakis}
\affiliation{Department of Materials Science and
Technology, University of Crete, P.O. Box 2208, 71003 Heraklion, Crete}
\author{Pantelis C. Kelires}
\affiliation{Department of Physics, University of Crete, P.O. Box 2208, 71003
Heraklion, Crete, Greece}
\affiliation{Department of Mechanical Engineering and Materials Science and
Engineering, Cyprus University of Technology, P.O. Box 50329, 3603 Limassol,
Cyprus}

\begin{abstract}
Ultra-nanocrystalline diamond is a polycrystalline material, having
crystalline diamond grains of sizes in the nanometer regime. We study the
structure and mechanical properties of this material as a function of the
average grain size, employing atomistic simulations. From the calculated
elastic constants and the estimated hardness, we observe softening of the
material as the size of its grains decreases. We attribute the observed
softening to the enhanced fraction of interfacial atoms as the average grain
size becomes smaller. We provide a fitting formula for the scaling of the
cohesive energy and bulk modulus with respect to the average grain size.  We
find that they both scale as quadratic polynomials of the inverse grain size.
Our formulae yield correct values for bulk diamond in the limit of large grain
sizes.
\end{abstract}

\date{\today}

\keywords{Nanocrystalline materials; carbon \& graphite; hardness. \vspace{6mm} \\ To be published in Acta Materialia. \vspace{6mm}}

\maketitle

\section{Introduction}

As most ordinary solids are polycrystalline, the dependence of mechanical and
other properties on the size of their grains is a question of fundamental
interest for materials science \cite{meyers06}. The hardness of
polycrystalline metals increases with decreasing grain size, in accordance to
the Hall-Petch law: their yield stress is a linear function of $d^{-n}$, where
$d$ is the average grain size and $n>0$.  Mechanical load beyond the elastic
regime is mostly undertaken by dislocations, and plastic deformation usually
involves dislocation motion.  By introducing more grain boundary area in the
material, when grains become smaller, the motion of dislocations is impeded
and dislocations tend to pile up near grain boundaries. This in turn yields
harder materials \cite{zhao03,bata04}.

On the other hand, when the grain size reaches the nanometer range, several
metals have been found to exhibit a so-called ``reverse Hall-Petch effect'',
and become softer at smaller grain sizes \cite{schiotz98,conrad00}. The
presence of dislocations no longer governs the mechanical response on this
regime. As the fraction of atoms residing on grain boundaries increases with
decreasing grain size, any external mechanical load will be primarily
undertaken by sliding along grain boundaries
\cite{schiotz98,swygenhoven99,yamakov04,shimokawa05,keblinski98}. The
interplay between these bulk- and grain-boundary-related mechanisms of
plasticity can result to an optimum size of the grains, for which the material
is strongest \cite{yip98,argon06}. This size is of the order of 10-15 nm for
Copper \cite{schiotz03}. Contrary to these studies for metals, very few
workers have addressed the dependence of mechanical properties on the grain
size for nanocrystalline semiconductors or insulators. Examples include a
theoretical observation of higher creep rates at smaller grain sizes for
nanocrystalline Si \cite{keblinski98} or an experimental observation of a
maximum strength for an intermediate grain size in BN nanocomposites
\cite{dubrovinskaia07}.

Several pioneering works have dealt with the mechanical properties of
nanocrystalline semiconductors and insulators, like Si
\cite{keblinski98,demkowicz07} and SiC \cite{szlufarska05}. In the present
work, we take one step further and investigate the dependence of the
mechanical properties on the average grain size in ultra-nanocrystalline
diamond (UNCD). C forms extremely strong and directional bonds, while, at the
same time, the availability of $sp^2$ and $sp^1$ hybridizations allows for the
presence of non-defective under-coordinated atoms. This results in lower
relative grain-boundary energies compared to other group-IV elements
\cite{keblinski98b}. It is then expected that the difference between the
strengths of inter- and intra-grain bonds should be larger in polycrystalline
diamond compared to other polycrystalline materials. For this reason, UNCD can
serve as a prototype for the mechanical properties of nanocrystalline
ceramics.

Ultra-nanocrystalline diamond (UNCD) is a polycrystalline carbon-based
material, having grains a few nanometers big \cite{gruen99}. It is a low-cost
material with a potential for a wide range of applications due to its unique
mechanical and electronic properties \cite{krauss01}; in addition, its
properties can be tailored with appropriate doping \cite{rovere06}. Despite
the strong directional C-C bonds, resulting in inhomogeneity at the atomic
scale, the material can be considered as isotropic at larger scales, as no
particular orientation for the grain boundaries in UNCD seems to be favoured
in the experiment \cite{gruen99}. Theoretical calculations show that the grain
boundary energies have a weak dependence on the orientation \cite{zapol01}, as
do the energies of the interfaces between amorphous C and diamond
\cite{kopidakis07}. Sizes of grains were found to have a broad distribution,
with an average of about 3 nm and most grains being between 2 and 5 nm
\cite{gruen99}. The properties of the material can be tailored by modifying
the dopand concentration or the preparation conditions
\cite{philip03,zshen06}.

\section{Computational method}

Theoretical modelling of the mechanical properties of UNCD has only been made
possible so far by employing either small clusters\cite{paci05} or infinite
rods\cite{shen06,angadi06}. Here, we present a fully three-dimensional,
computer-generated atomistic model of UNCD, having grains of different sizes
separated by random grain boundaries. The simulations were performed using a
continuous-space Monte Carlo method. We employ the many-body potential of
Tersoff\cite{tersoff88}. The choice of the empirical potential is important;
for example, different potentials give different responces in Si under large
pressures \cite{godet04}. The Tersoff potential used here provides a very good
description of the structure and energetics for a wide range of carbon-based
materials\cite{kelires94,kelires00}. This method, although considerably
demanding computationally, allows for great statistical accuracy, as it is
possible to have samples at full thermodynamic equilibrium.  Such an accuracy
is neccesary due to the many different possible hybridizations of C atoms.  In
addition, a reliable calculation of mechanical properties requires a fully
relaxed structure. Residual stress can affect the mechanical properties,
especially in nanocrystalline materials.  The supercells we use contain up to
about 120000 C atoms.

\begin{figure}
\begin{center}
\includegraphics[width=0.7\columnwidth]{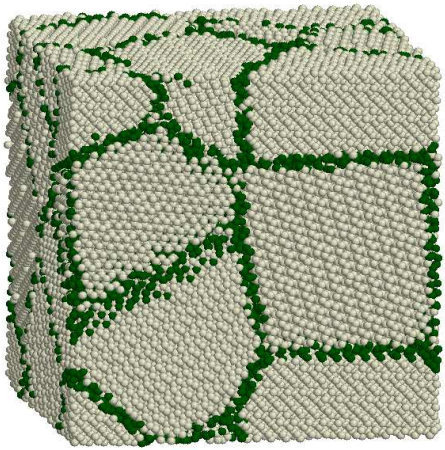}
\end{center}
\caption{One of the UNCD samples used in the simulations. The
system contains eight grains, with average grain diameter equal to 4.4 nm and
contains 116941 atoms in total. Atoms that have less than four neighbors, or
deviate significantly from the tetrahedral geometry, are shown in darker
color.}
\label{fig:structure}
\end{figure}

We model ultra-nanocrystalline diamond (UNCD) by a periodic repetition of
cubic supercells that consist of eight different regions (grains) each. The
number of grains in the unit cell guarantees the absence of artificial
interactions between a grain and its periodic images. The grains have random
shapes and sizes, and are filled with atoms in a randomly oriented diamond
structure. The method we use is identical to the method used by Schiotz and
co-workers to model nanocrystalline copper \cite{schiotz98,schiotz03}. To
achieve a fully equilibrated structure for each grain size, we perform four
steps: first, the structure is compressed and equilibrated at constant volume
at 300 K, in order to eliminate large void regions near some grain boundaries,
that are an artifact of the randomly generated structure. In the second step,
we anneal the system at 800 K allowing volume relaxation and quench down to
300 K. Third, we anneal once more, at 1200 K this time, in order to ensure
full equilibration. Fourth, we fully relax the structure at 300 K allowing for
changes in both volume and shape of the unit cell.

\section{Results}

\subsection{Structure}
The relaxed structure for a typical sample is shown in
Fig. \ref{fig:structure}. The grain boundaries are a few atomic diameters
wide, in accordance with experiments showing widths of 0.2-0.5 nm
\cite{gruen99}. Atoms at the grain boundaries are either three-fold
coordinated or form bonds at different lengths or angles from those observed
in diamond. The structural and elastic properties for characteristic samples
are summarised in Table I. The fraction of the three-fold atoms in the samples
is about 1/10 for grain sizes between 3.5 and 4.5 nm; in experiment, it was
observed that the fraction of atoms residing at grain boundaries is close to
10\% for similar crystallite sizes \cite{gruen99}.

\begin{table}
\caption{Properties of characteristic UNCD samples at 300K: average grain size
($d$, in nm), number of atoms in the simulation cell ($N$), percentage of
three-fold atoms in the cell ($N_3$, at \%), mass density ($\rho$, in g/cc),
cohesive energy ($E_{coh}$, in eV per atom), bulk modulus ($B$, in GPa),
Young's modulus ($E$, in GPa) and shear modulus ($G$, in GPa). For comparison,
the corresponding values for single-crystal diamond, calculated with the same
method, are shown in the last line (from Ref.\cite{kelires94}).}
\begin{center}
\begin{tabular}{rrrrrrrrr} \hline\hline
$d$  & $N$    & $N_3$& $\rho$&$E_{coh}$& $B$ & $E$ &$G$\\ \hline
1.94 & 9,149  & 33.6 & 3.14 & -7.00 & 303 & 719 &  325 \\
2.43 & 18,528 & 26.4 & 3.22 & -7.06 & 323 & 808 &  373 \\  
2.93 & 33,170 & 20.5 & 3.28 & -7.11 & 342 & 891 &  418 \\
3.44 & 53,494 & 12.0 & 3.30 & -7.10 & 363 & 939 &  439 \\
3.92 & 81,561 & 10.2 & 3.37 & -7.14 & 372 & 963 &  451 \\
4.41 &116,941 & 9.12 & 3.40 & -7.15 & 384 & 987 &  461 \\
$\infty$&$\infty$&0.00&3.51 & -7.33 & 443 &1,066&  485 \\
\hline\hline
\end{tabular}
\end{center}
\end{table}

\begin{figure*}
\includegraphics[width=\textwidth,clip=true]{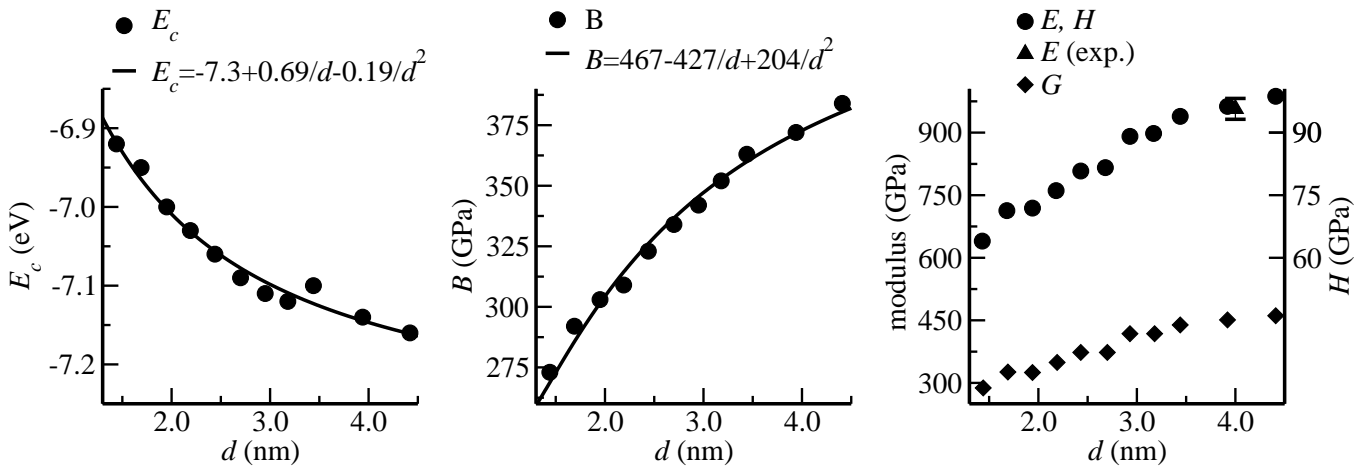}
\caption{Cohesive energy (left panel), bulk modulus (center panel), Young's
and shear moduli and estimated hardness (right) of UNCD versus the average
grain size. The solid lines in the left and central panel are fits to the
data. The function used assumes different compressibilities for atoms in the
bulk, grain boundaries or edges.  In the right panel, the experimental value
from Ref. \cite{espinosa06} is also shown.}
\label{fig:by_vs_d}
\end{figure*}

\subsection{Elastic moduli and estimated hardness}
The bulk modulus ($B$) and Young's modulus ($E$) are calculated by applying
small hydrostatic and uniaxial deformations to the material, respectively. We
find the moduli by fitting a parabola to the thus obtained energy versus
strain data. In the case of Young's modulus, we apply load along three
mutually perpendicular directions, and take the average. Bulk moduli of UNCD
samples are plotted against the average grain size in
Fig. \ref{fig:by_vs_d}. They constantly decrease with decreasing grain size,
offering a first evidence for softening of the material at small grain
sizes. The same behavior is observed through the Young's and shear moduli that
are shown in Fig. \ref{fig:by_vs_d}. \footnote{The shear moduli, $G$, are
calculated from the values of $B$ and $E$. For an isotropic material there are
only two independent elastic constants; the formula connecting $B$, $E$ and
$G$ is $\displaystyle G=\frac{3BE}{9B-E}$.}  For the experimentally relevant
range of grain sizes, between 3 to 5 nm, UNCD samples are found to have very
high elastic moduli, only slightly lower than those of diamond, placing thus
UNCD into the family of super-hard materials \cite{kaner05}. The Young's
modulus of high-purity UNCD was measured to be 957$\pm$25 GPa
\cite{espinosa06}. This is in excellent agreement with our result of 963 GPa
for a grain size of 3.9 nm.

All elastic moduli are found to decrease with decreasing average grain size,
indicating softening of the material. This suggests that, in this range of
grain sizes, the hardness of the material should also drop with decreasing
grain size. Indeed, the hardness of many materials is proportional to the
Young's or shear modulus \cite{brazhkin02}. For nanocrystalline materials,
extended defects, like cracks, cannot exist as their lengths are in the
micrometer range \cite{bobrovnitchii07}, rendering the elastic constants good
descriptors of hardness.  In particular, the hardness of carbon-based
materials has been found to be between 10\% and 16\% of the Young's modulus
\cite{robertson02}. This allows us to make a rough estimate for the hardness
of UNCD, as being roughly one tenth of its Young's modulus. We use this
approximation to plot the estimated hardness of UNCD as a function of grain
size in Fig. \ref{fig:by_vs_d}.

\section{Discussion}

Our results, together with observations for nanocrystalline copper
\cite{schiotz98}, silicon \cite{keblinski98}, and boron nitride
\cite{dubrovinskaia07}, show that softening at small grain sizes can occur in
various different nanocrystalline materials. This might be understood
considering the relative numbers of atoms near boundaries and in the bulk of
grains. These numbers become comparable for grains in the nanometer regime, no
mater what the particular chemical composition of the material is. Atoms at
grain faces, edges or vortexes, as well as atoms near other discontinuities,
will naturally form bonds that are weaker than those formed by atoms in the
bulk. Such weaker bonds will then bend or stretch with greater ease, compared
to the bonds in the crystalline region. This explains the softening of
polycrystalline solids when the grain size is at the nanometer range. For much
larger grain sizes, the number of grain-boundary atoms will be negligible
compared to the number of bulk atoms; in this regime, the behavior of the
material under mechanical load will be characterised mostly by bulk defects,
such as dislocations.

\begin{figure}
\begin{center}
\includegraphics[width=0.7\columnwidth]{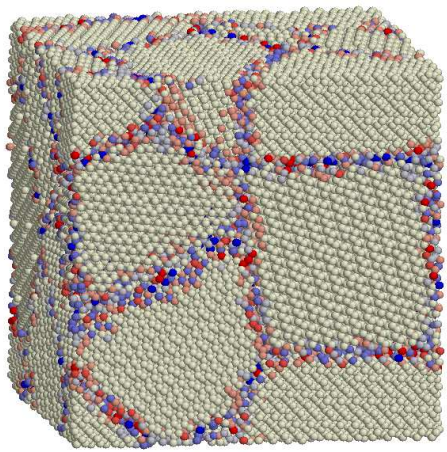}
\end{center}
\caption{The UNCD sample shown in Fig. \ref{fig:structure}, with atoms being
colored according to their local bulk moduli: atoms having bulk moduli within
10\% of the average value are colored gray. Blue and red color indicate
deviation from the average value towards higher or lower values,
respectively. Darker colors indicate higher absolute value of the deviation. }
\label{fig:local_b}
\end{figure}

To make this picture quantitative, let us divide the atoms in the
polycrystalline material into three categories: (a) Atoms deep inside the
grains, forming bonds that are roughly identical to those in the
single-crystal material. Their number is proportional to $d^3$, where $d$ is
the average grain size. (b) Atoms near the grain boundary; these behave
similarly to surface or interface atoms. Their number is proportional to
$d^2$. (c) Atoms near grain boundary edges; these are similar to kink surface
atoms, or atoms near dislocation cores. Their number is proportional to
$d$. Of course, there will be other types of atoms, like vertex atoms or atoms
near topological defects, but their number will be much smaller than the
numbers of atoms falling in one of the aforementioned categories.  The
cohesive energy of the solid will be the sum of the energies of the three
different atom types, multiplied by their respective numbers, and divided by
the total number of atoms, which is proportional to $d^3$. Therefore, the
cohesive energy should be described by a function of the form
$E_{coh}=E_0+a/d+b/d^2$, where $a$, and $b$ are constants, and $E_0$ is the
cohesive energy of the monocrystalline solid. Indeed, such a function fits our
data perfectly, the rms error being less than 0.5\%. Moreover, $E_0$ is found
to be -7.31 eV, very close to the calculated cohesive free energy of diamond
at 300 K which is -7.33 eV. As $B$ is proportional to the second derivative of
the total energy with respect to the system volume, it can also be decomposed
into contributions from bulk, interface and vertex atoms. As shown in
Fig. \ref{fig:by_vs_d}, a parametrisation of the form of a quadratic equation
in 1/$d$ fits very nicely the results of the simulation. The constant value,
467 GPa, corresponding to the ideal monocrystalline solid, is only 5\% off the
calculated value for diamond. Such a decomposition of the total bulk modulus
to a sum of atomic-level moduli has been used previously, in order to
investigate the rigidity of amorphous carbon
\cite{kelires00}. Fig. \ref{fig:local_b}, showing the local bulk moduli in the
sample, demonstrates the existence of atoms near the grain boundaries having
bulk moduli well away from the average value. It is these atoms that
contribute to the terms proportional to $1/d$ and $1/d^2$ in the function that
fits the simulation data for the bulk modulus.

In this particular example of a nanocrystalline material, under-coordinated
atoms may not necessarily be considered to be defects. Carbon atoms are known
to exist in several hybridizations, and bonds between two-fold coordinated
$sp^1$ atoms or between three-fold coordinated $sp^2$ atoms are usually
stronger than bonds between four-fold coordinated $sp^3$ atoms. For example,
in amorphous carbon, where all these hybridizations co-exist, it is the
failure of bonds between $sp^3$ atoms that governs the fracture of the
material \cite{fyta06}. The ability of carbon to have $sp^2$ atoms renders
grain boundaries in UNCD to be extremely stable \cite{keblinski99}. On the
other hand, bonds between $sp^2$ or $sp^1$ atoms are stabilised by
$p$-bonding, which is very sensitive to the geometry. As a consequence, local
bulk moduli of $sp^1$ and $sp^2$ atoms are found to be significantly lower
than the bulk moduli of $sp^3$ atoms \cite{kelires00}. This is demonstrated by
the lower bulk modulus of $sp^2$-rich amorphous carbon compared to $sp^3$-rich
one \cite{mathioudakis04}. Here, a similar analysis of some UNCD samples
reveals that the average bulk modulus of the three-fold atoms ($\sim$ 250 GPa)
is much lower than the average bulk modulus of the four-fold atoms ($\sim$ 420
GPa), which, in turn, predominates the bulk modulus of the sample.

\section{Conclusions}

Using ultra-nanocrystalline diamond (UNCD) as a prototype for a
polycrystalline covalent solid with grains at the nanometer regime, we have
observed softening of the material as the grain size decreases, in analogy
with the reverse Hall-Petch effect observed in nanocrystalline metals.  The
effect is attributed to the increasing fraction of grain-boundary atoms as the
grain size decreasing. A simple quadratic form in $1/d$, where $d$ is the
average grain size, suffices to fit the results for both cohesive energy and
bulk modulus, while yields the correct values for bulk diamond for large grain
sizes. The measured Young's modulus of UNCD is reproduced well by the
simulations. Our results provide further evidence that softening at low grain
sizes can occur in various kinds of nanocrystalline solids.

\vspace{1.5cm}

{\parindent 0cm \bf \Large Acknowledgement}

\vspace{0.5cm}

This work is supported by a grant from the Ministry of National Education and
Religious Affairs of Greece through the action ``$\mathrm{E\Pi EAEK}$''
(programme ``$\mathrm{\Pi Y\Theta A\Gamma OPA\Sigma}$.'')

\end{document}